\documentclass[conference]{IEEEtran}
\usepackage{myStyleIEEE}
\usepackage{ushort}
\IEEEoverridecommandlockouts

\def\R{\mathbb{R}}
\def\C{\mathbb{C}}

\usepackage{siunitx}
\AtBeginDocument{\DeclareSIUnit{\MWh}{MWh}}
\AtBeginDocument{\DeclareSIUnit{\kWh}{kWh}}
\usepackage{amsmath}
\usepackage{mathtools}
\usepackage{isomath}

\usepackage[noadjust]{cite}

\begin{document}

\title{Fast Mapping of Flexibility Regions at TSO-DSO Interfaces under Uncertainty}

\renewcommand{\theenumi}{\alph{enumi}}

\author{
\IEEEauthorblockN{Alice Patig\IEEEauthorrefmark{1}\IEEEauthorrefmark{3}, Ognjen Stanojev\IEEEauthorrefmark{1}, Petros Aristidou\IEEEauthorrefmark{2}, Aristides Kiprakis\IEEEauthorrefmark{3}, Gabriela Hug\IEEEauthorrefmark{1}}%
\IEEEauthorblockA{\IEEEauthorrefmark{1} EEH - Power Systems Laboratory, ETH Zurich, Switzerland } %
\IEEEauthorblockA{\IEEEauthorrefmark{2} Department of Electrical Engineering, Computer Engineering and Informatics, Cyprus University of Technology, Cyprus} %
\IEEEauthorblockA{\IEEEauthorrefmark{3} Institute for Energy Systems, School of Engineering, University of Edinburgh, United Kingdom} %
Emails: \{stanojev, hug\}@eeh.ee.ethz.ch, alipatig@gmail.com,  petros.aristidou@cut.ac.cy, aristides.kiprakis@ed.ac.uk
%\vspace{-0.45cm}
\thanks{Research supported by NCCR Automation, a National Centre of Competence in Research, funded by the Swiss National Science Foundation (grant number 180545). The authors would further like to thank Dr. Uros Markovic and Dr. Dmitry Shchetinin for fruitful discussions.}
}

\maketitle
\IEEEpeerreviewmaketitle

%ABSTRACT
\begin{abstract}
The gradual decommissioning of fossil fuel-driven power plants, that traditionally provide most operational flexibility in power systems, has led to more frequent grid stability issues. To compensate for the lack of flexible resources, Distributed Energy Resources (DERs) in distribution networks can be employed. To facilitate the use of DERs, the aggregated flexibility in a distribution grid is commonly represented on a $PQ$-plane displaying the feasible active and reactive power exchange with the upstream grid. This paper proposes a fast feasible operating region mapping mechanism that utilizes a linear power flow approximation in combination with linearized generator, current, and voltage constraints to construct a high-dimensional polyhedral feasible set of DER power injections. The obtained polytope is projected onto the $PQ$-plane using Fourier-Motzkin Elimination to represent the aggregate network flexibility. Additionally, uncertainty in DER generation is addressed using chance-constraints. Our analysis of a modified 33-bus IEEE test system demonstrates that the proposed method obtains more accurate approximations than former geometric methods and is ten times faster than the tested optimization-based method.
\end{abstract}

%INDEX TERMS
\begin{IEEEkeywords}
active distribution networks, distributed energy resources, TSO-DSO coordination, flexibility computation
\end{IEEEkeywords}

\section{Introduction} \label{sec:intro}
Fossil fuel-driven power plants traditionally provide most operational flexibility in power systems. However, due to their contribution to climate change through greenhouse gas emissions, they are increasingly replaced by renewable energy sources. The resulting power systems are subject to more frequent grid congestion and stability issues~\cite{Milano2018}. Hence, new sources of flexibility are needed for congestion management, voltage and frequency control. At the same time, the prices for solar PhotoVoltaic (PV) and Battery Energy Storage Systems (BESS) are decreasing, and the research effort in smart grid control schemes for better Distributed Energy Resource (DER) coordination is increasing, leading to growing DER investments in distribution grids \cite{Hatziargyriou2017}. The coordinated dispatch of DERs located in distribution grids can be a solution to provide the needed flexibility and support the transmission system \cite{Hatziargyriou2017}. 
%With these developments in mind, it is clear that the realization of a more vital Transmission System Operator - Distribution System Operator (TSO-DSO) interaction is becoming imperative for the safe operation of future power grids.

%With these developments in mind, it is clear that the realization of a stronger Transmission System Operator (TSO) - Distribution System Operator (DSO) interaction is becoming an imperative for the safe grid operation \cite{EUCR}.
%Therefore, multitudes of heterogeneous DERs in Active Distribution Networks (ADNs) can be aggregated and used as a new flexibility source to the transmission system~\cite{EUCR}. 

To help system operators access the flexibility in Active Distribution Grids (ADNs) more easily, information on the feasible dispatch points of the ADN at the Transmission System Operator (TSO) - Distribution System Operator (DSO) interface is needed \cite{CapitanescuEPSR2018}. The set of all feasible dispatch points represents the aggregated flexibility of an ADN and forms a region in the $PQ$-plane, where $P$ is the active and $Q$ the reactive power exchange between the distribution and transmission grid. We will refer to this set as the \textit{Feasible Operating Region} (FOR) based on its definitions given by \cite{Riaz2019}. Constructing this feasible region is challenging, as the capability curves of the DERs within the ADN, the network constraints, and the uncertainty inherent to most DERs must be considered and rapid computation is required in real-time.
% Moreover, real-time operation requires rapid computation of the region. 

%Hence, the goal of this work is to display the aggregated flexibility of an ADN on the $PQ$-plane, where $P$ is the active and $Q$ the reactive power exchange between the distribution and transmission grid. 
%In other words, we are constructing the set of technically feasible operating points of the ADN, similarly to a generator’s capability curve, based on network and aggregated DER constraints. We will refer to this set as the \textit{Feasible Operating Region} (FOR) based on its definitions given by \cite{Riaz2019}.

Distribution network FOR mapping was first undertaken using a Monte Carlo method in \cite{HelenoPowerTech2015}. This stochastic approach is computationally intensive and defines the edge-regions with low accuracy. Hence, optimization-based approaches were developed in which the boundary of the FOR is point-wisely computed using AC Optimal Power Flow (OPF) \cite{Silva2018, CapitanescuEPSR2018, Sarstedt2021}. However, the methods' computational complexity grows exponentially with system size and is non-negligible even for small networks. Geometric approaches have attempted faster FOR mapping~\cite{Zhao20174721,Fortenbacher2020}. In \cite{Zhao20174721}, the individual DER capability curves are aggregated using the Minkowski Sum method to obtain the FOR, thus neglecting the network constraints. The polyhedral method in \cite{Fortenbacher2020} includes generator and network constraints which results in a high-dimensional FOR in terms of DER injections. The injections are then related to $P$ and $Q$ as fixed fractions using Generation Shift Keys (GSKs). Hence, this method only reveals scenarios in which DERs operate in fixed proportion to each other, thus hiding large parts of the true FOR and being highly dependent on the chosen GSK scheme. Furthermore, the DER uncertainty is neglected.

In this paper, the aim is to develop a novel FOR mapping approach that is faster than the Monte Carlo-based and optimization-based methods but more accurate than the earlier fast geometric methods. This novel approach seeks to be more suitable for real-time use by system operators. Similarly to \cite{Fortenbacher2020}, we establish linear network and aggregated generator constraints. However, we avoid the use of GSKs by establishing the relationship among DER injections and $P$ and $Q$ flexibility using a single Backward/Forward Sweep (BFS) power flow iteration. The high-dimensional FOR in terms of DER injections is then projected onto the $PQ$-plane using Fourier-Motzkin Elimination (FME). Uncertainty is addressed by chance-constraining uncertain DER generation such that the level of risk in the FOR mapping can be adjusted. %Only few other methods take uncertainty into account \cite{Kalantar-Neyestanaki20202464}. 

The remainder of the paper is structured as follows. In Section~\ref{sec:flex_modeling}, the network model is introduced and the notion of distribution grid flexibility is defined. Section~\ref{sec:FME} details how the BFS power flow equations and the FME projection mechanism are used to rapidly map the FOR onto the $PQ$-plane. Section~\ref{sec:CCO} discusses two approaches to handle uncertainty in DER generation within the proposed FOR mapping method. Section~\ref{sec:res} assesses the performance of the proposed approach in a case study and compares it to existing methods, while Section~\ref{sec:concl} concludes the paper and discusses future work. %avenues for

\textit{Notation}: We denote the sets of natural, real and complex numbers by $\N,\,\R$ and $\C$, and define $\R_{\geq a}\coloneqq\{x\in\R\mid x\geq a\}$. The real part of a complex number $z\in\C$ is denoted by $\mathfrak{R}(z)$ and the imaginary part by $\mathfrak{I}(z)$. The complex imaginary unit is denoted by $j\coloneqq\sqrt{-1}$. For column vectors $x\in\R^n$ and $y\in\R^m$ we use $(x,y)\coloneqq [x^\top,y^\top]^\top\in\R^{n+m}$ to denote a stacked vector, and $\|x\|$ denotes the Eucledian norm. Finally, $\mathbbl{1}_n$ represents a column vector of ones of length $n$.

\section{Modeling of Distribution Grid Flexibility} \label{sec:flex_modeling}
\subsection{Network model}
This work assumes a balanced radial distribution network represented by a tree graph $\mathcal{G}=(\mathcal{N},\mathcal{E})$, where $\mathcal{N} \coloneqq \{0,1,\dots,n\}$ denotes the set of network nodes including the substation node $0$, and $\mathcal{E} \subseteq \mathcal{N}\times\mathcal{N}$ represents the set of $n$ network branches. A single Point of Common Coupling (PCC) between transmission and distribution grid at node 0 is assumed. One renewable energy source, one storage, and one conventional generator type were considered for proof of concept. Let $\mathcal{P}\subseteq \mathcal{N}$ be the subset of nodes which host PV units, $\mathcal{B}\subseteq \mathcal{P}\subseteq \mathcal{N}$ the subset of nodes with BESS, and $\mathcal{D}\subseteq \mathcal{N}$ the subset of nodes which host Diesel Generators (DGs). Storage is assumed to be installed only at nodes with PV, as those participants who install PV are assumed to be more likely to invest in storage. The set of DERs in the ADN is thus $\mathcal{C} \coloneqq \mathcal{P} \cup \mathcal{B} \cup \mathcal{D}$, with cardinality $n_c\coloneqq|\mathcal{C}|$.

For every node $i\in\mathcal{N}$, let $V_i\in\C$ denote the complex voltage, and let $P_{i}^\mathrm{inj}\in\R$ and $Q_{i}^\mathrm{inj}\in\R$ represent the active and reactive power injections. For each branch $(i,j)\in\mathcal{E}$, let $I_{ij}\in\C$ represent the corresponding branch current. The vectors of the introduced quantities $V\in\C^n,\,P^\mathrm{inj}\in\R^n,\,Q^\mathrm{inj}\in\R^n$ and $I\in\C^n$ are obtained by stacking the appropriate bus and branch variables, e.g., $V = (V_1,V_2,\dots,V_n)$.  The distribution grid is typically modeled using the power flow equations, which allow to define an input-output map $\ell(\cdot)$ relating the previously introduced physical quantities, as follows:  
\begin{equation} \label{eq:power_flow}
    (V,I) = \ell(P^\mathrm{inj},Q^\mathrm{inj}).
\end{equation}
While the input-output map cannot be written analytically in a compact closed form, the implicit function theorem~\cite{implicit_function_theorem} guarantees the local existence of a continuous differentiable map $\ell(\cdot)$. A linear approximation of $\ell(\cdot)$ around a specific operating point is introduced in a subsequent section.

\subsection{Distribution Grid Flexibility as a Feasible Set}
\begin{figure}[!t]
    \centering
    % \vspace{-0.35cm}
    \includegraphics[scale=0.925]{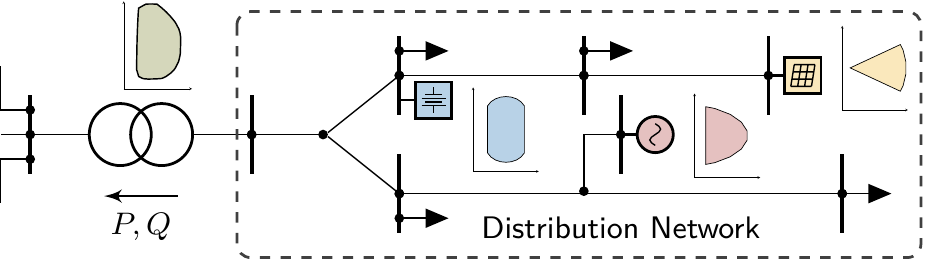}
    \caption{An exemplary distribution grid featuring three DER units: a BESS (in blue), a DG (in red), and a PV system (in yellow). The two-axis coordinate systems are active-reactive power planes representing the capability curve of the corresponding unit. The aggregated $PQ$ flexibility at the TSO-DSO interface is presented by the green region.}
    \label{fig:FOR_mapping_illustration}
     \vspace{-0.35cm}
\end{figure}

As mentioned earlier, the goal is to determine the set of feasible operating points of the ADN as seen at the PCC, such that the TSO can use this information to deploy the aggregated flexibility of DERs for ancillary services. To do so, we can use the concept of a \textit{feasible set} from optimization theory, which is the set of all possible values that the decision variables can take while satisfying the problem's constraints. In the problem at hand, we assume that the system operators are the decision makers who control the active and reactive power outputs $(p_i,q_i)\in\R^2$ of each DER unit $i\in\mathcal{C}$. Assuming lossless interfaces of DER units to the grid, we can map the DER generation variables to the power injection vector as:
\begin{equation}\label{eq:cmpt_injections}
    \begin{bmatrix}
       P^\mathrm{inj} \\ Q^\mathrm{inj}
    \end{bmatrix} = C_g 
    \begin{bmatrix}
        p \\ q  
    \end{bmatrix}-L,
\end{equation}
where $p\coloneqq (p_1,\dots,p_{n_c})$ and $q\coloneqq (q_1, \dots, q_{n_c})$ are the vectors containing individual DER power generation elements, $L\in\R^{2n}$ is a vector of active and reactive power demands, and $C_g\in\{0,1\}^{2n\times2n_c}$ is a binary matrix derived from the DER placement. As discussed in the following, constraints on the decision variables $(p,q)$ are imposed by the capability curves of the individual units as well as the voltage and current limits of the network. An exemplary grid with indicated feasible regions of DER units and the ADN FOR is given in Fig.~\ref{fig:FOR_mapping_illustration}.
%An exemplary distribution grid with indicated feasible regions of DER units, considering their individual capability curves, and the aggregated $PQ$ region (considering the network limits as well) is depicted in Fig.~\ref{fig:FOR_mapping_illustration}.
%\subsubsection{Generator Constraints - Individual Capability Curves} \label{sec:gen_constr} \cite{pf}

The active and reactive power outputs of DER units need to comply with their electro-mechanical and thermal limits represented by the so-called capability curves. Typically, these are defined as sets of allowable setpoints $(p_{i},q_{i}) \in\mathcal{X}_{i}\subseteq\R^2$ for each unit $i\in\mathcal{C}$. The capability curve of the PV generators $i\in\mathcal{P}$ is defined as the following set: 
\begin{equation*}
\begin{split}
    \mathcal{X}_i \coloneqq \{(p_{i},q_{i}): 0\leq p_{i} \leq p^\mathrm{pv}_{i}, p_{i}^2+q_{i}^2\leq s^\mathrm{max}_i,  \cos{\phi}\geq a\},
\end{split}
\end{equation*}
where $s^\mathrm{max}_{i}\in\R_{\geq0}$ denotes the squared apparent power limit and $p^\mathrm{pv}_{i}\in\R_{\geq0}$ represents the irradiation dependent active power upper limit, while
the lower limit is zero due to the possibility of curtailment. Furthermore, power quality standards require the power
factor $\cos{\phi}$ to be greater than a prescribed value $a\in[0,1]$.
The BESS capability curve is assumed to allow four-quadrant operation for each unit $i\in\mathcal{B}$:
\begin{equation*}
    \mathcal{X}_i \coloneqq \{(p_{i},q_{i}): -p^\mathrm{max}_{i} \leq p_{i} \leq p^\mathrm{max}_{i}, p_{i}^2+q_{i}^2\leq s^\mathrm{max}_i\},
\end{equation*}
with $p^\mathrm{max}_{i}\in\R_{\geq0}$ denoting the maximum charging/discharging power. Storage energy capacity limits are not considered as the FOR computation is conducted for a particular time instant. Finally, operation of each DG unit $i\in\mathcal{D}$ is limited by stator current and low load limits, which impose maximum apparent power output $s_i^\mathrm{max}\in\R_{\geq0}$ and minimum active power generation $p_i^\mathrm{min}\in\R_{\geq0}$ constraints, as follows: 
\begin{equation*}
    \mathcal{X}_i \coloneqq \{(p_{i},q_{i}): p^\mathrm{min}_i \leq p_{i},\,\, p_{i}^2+q_{i}^2\leq s^\mathrm{max}_i\}.
\end{equation*}

%\subsubsection{Grid Constraints - Flow and Voltage Deviation Limits} \label{sec:hnet}
Generator capabilities do not pose the only constraints when attempting to extract the flexibility available in an ADN. Network constraints, in the form of voltage deviations and line current limits, must be considered to respect the power quality standards and thermal ratings of assets. Finally, the feasible set of DER active and reactive power injections representing the distribution grid flexibility can be formulated as:
\begin{equation} \label{eq:feasible_space}
\begin{split}
    \mathcal{F} \coloneqq \big\{&V^\mathrm{min}\leq \|V\| \leq V^\mathrm{max},\,\|I\| \leq I^\mathrm{max},\\&(V,I) = \ell(C_g(p,q)-L),  (p_i,q_i)\in\mathcal{X}_i,\forall i\in\mathcal{C}\big\},
\end{split}
\end{equation}
where $V^\mathrm{min}\in\R_{\geq0}^{n}$ and $V^\mathrm{max}\in\R_{\geq0}^{n}$ denote the vectors of minimum and maximum nodal voltage magnitudes, and $I^\mathrm{max}\in\R^n_{\geq0}$ is the vector of thermal line limits.

\section{Fast Flexibility Mapping using Fourier-Motzkin Elimination} \label{sec:FME}

In computational geometry, a set that is defined by linear constraints can be displayed as a polytope, which is a geometric object with any number of flat sides. Hence, if we linearise the constraints in \eqref{eq:feasible_space}, we can display the feasible set of DER generation variables $(p,q)$ as a $2n_c$-dimensional polytope, on a graph with $n_c$ $p$-axes and $n_c$ $q$-axes, showing ADN's operating scenarios in terms of active and reactive power generation of each DER which neither violate individual generator constraints nor collectively any network constraints. However, the goal is to display the aggregated flexibility of the ADN as a function of the active and reactive power flow at the PCC, $(P,Q)\in\R^2$. To this end, a linear power flow approximation is introduced in Section~\ref{subsec:BFSdecomposed} to relate $(p,q)$ to $(P,Q)$. The polytope in $2n_c$ dimensions can then be \textit{projected} onto the $2$-dimensional $PQ$-plane. Algebraically, the projection is done using the Fourier-Motzkin Elimination \cite{fouriermotzkin} by reducing the number of dimensions iteratively while preserving each variable's effect on the remaining set.  

\subsection{Linear Power Flow Model using the BFS Method} \label{subsec:BFSdecomposed}
We employ the Backward/Forward Sweep method~\cite{Teng2003} to linearly approximate $\ell(\cdot)$ in \eqref{eq:power_flow} around an operating point. The method starts with a linearization step, where the nodal current injections $I^\mathrm{inj}\in\C^{n}$ are computed based on the active and reactive power injections $(P^\mathrm{inj},Q^\mathrm{inj})$ and a precomputed voltage operating point $\bar{V}\in\C^n$. Then, the backward sweep is performed, where the branch currents $I$ are calculated using these current injections $I^\mathrm{inj}$. Finally, in the forward sweep, nodal voltages $V$ are updated by computing voltage drops based on the current injections and subtracting them from a vector of slack bus voltage magnitudes $V_{s}\in\C$. A single iteration of the BFS method can thus be summarized as:
\begin{align}
    I^\mathrm{inj} &= PICI \cdot (P^\mathrm{inj}, Q^\mathrm{inj}), \label{eq:bfs_inj_model}  \\ 
    I &= BIBC \cdot I^\mathrm{inj},  \label{eq:bfs_branch_model}\\
    V &= \mathbbl{1}_nV_{s} - DLF\cdot I^\mathrm{inj},    \label{eq:bfs_voltage_model} 
\end{align}
where $PICI\coloneqq[\mathrm{diag}(\bar{V}^*)^{-1}\,\,\mathrm{diag}(j\bar{V}^*)^{-1}]$, $BIBC \in \{0,1\}^{n\times n}$ captures the network topology, $BCBV \in\C^{n\times n}$ contains network line impedances. The power injections $(P^\mathrm{inj},Q^\mathrm{inj})$ are related to DER outputs $(p,q)$ using \eqref{eq:cmpt_injections}. The BFS can be run iteratively until convergence, with the newest voltage updates used in the linearization step in every iteration. However, a single BFS iteration was previously shown to be a close approximation of the nonlinear power flow~\cite{Ognjen}, and is thus used here as a linear power flow model.    

\subsection{Flexibility around an Operating Point: Decomposing the BFS using Superimposed Circuits}
To gain insights into the flexibility of the network around an operating point, we deploy the superposition principle for linear circuits \cite{stanojev2022multiple}. It is customary to assume that the initial state of the network is known (e.g., via state estimation techniques), represented by initial voltage $V_\mathrm{init}\in\C^n$, branch current $I_\mathrm{init}\in\C^n$ and DER injection $(p_\mathrm{init},q_\mathrm{init})\in\R^{2n}$ values. The superposition principle states that the network response to the input changes, i.e., DER injections, can be computed by superimposing the response resulting from the input changes on the initial operating point. More formally, let us decompose each DER's power output into the DER's initial operating point, denoted by '$\mathrm{init}$', and the deviation around the setpoint driven by DER injection changes, denoted by '$\Delta$'. Similarly, we decompose bus voltages and branch currents as these will be subject to constraints:
{\scalebox{1}{
\begin{minipage}{.45\linewidth}
\vspace{-0.05cm}
\begin{subequations} \label{eq:decomp_pq}
\begin{align}
 p &= p_\mathrm{init} + \Delta p,\label{eq:decomp_p} \\
 q &= q_\mathrm{init} + \Delta q,\label{eq:decomp_q} 
\end{align}
\end{subequations}
\end{minipage}%
\begin{minipage}{.45\linewidth}
\vspace{-0.05cm}
\begin{subequations} \label{eq:decomp_vi}
\begin{align} 
    V &= V_\mathrm{init} + \Delta V, \label{eq:decomp_v} \\
    I &= I_\mathrm{init} + \Delta I. \label{eq:decomp_i}
\end{align}
\end{subequations}
\end{minipage}
}}\vspace{0.3cm}

\noindent By substituting \eqref{eq:decomp_vi} into \eqref{eq:bfs_branch_model} and \eqref{eq:bfs_voltage_model}, and \eqref{eq:decomp_pq} into \eqref{eq:cmpt_injections}, the initial operating point values cancel out resulting in:
\begin{equation}  \label{eq:BFS_final}
     \begin{bmatrix}
        \Delta I\\
        \Delta V
     \end{bmatrix} = 
     \begin{bmatrix}
        BIBC \\
        -DLF
     \end{bmatrix}  \cdot PICI \cdot C_g \cdot
            \begin{bmatrix}
               \Delta p \\
               \Delta q
            \end{bmatrix}.
\end{equation}

Furthermore, to project the feasible set in $2n_c$ dimensions onto the 2-dimensional $PQ$-plane, the algebraic relationship between $(\Delta p,\Delta q)$ and $(\Delta P,\Delta Q)$ must be established. To this end, let $\Delta I_0\in\C$ be the change in current flowing across the TSO-DSO interface, expressed using \eqref{eq:BFS_final} as follows:
\begin{equation}
        \Delta I_0 = 
        BIBC_{1} \cdot PICI \cdot C_g \cdot
            \begin{bmatrix}
               \Delta p \\
               \Delta q
            \end{bmatrix}, \label{eq:BFS_final-zero}
\end{equation}
where $\mathrm{BIBC}_{1}\in \{0,1\}^{1\times n}$ is the first row of the $BIBC$ matrix. Subsequently, active and reactive power changes at the TSO-DSO interface $(\Delta P,\Delta Q)$ can be computed from $\Delta I_0$ and the nodal voltage at the slack bus as follows:
\begin{equation}
    \Delta P = \mathfrak{R}(V_{s} \Delta I_0^*), \quad \Delta Q = \mathfrak{I}(V_{s}\Delta I_0^*).
\end{equation}
The decomposition hence obtains linear closed-form expressions that relate the changes in branch currents, bus voltages and PCC power flow directly to the changes in DER output.
%, enabling us to enforce linear voltage and current constraints. %Feasible deviations of DER outputs from the initial values represent the network flexibility around the initial operating point.

\subsection{Linear Current, Voltage and Generator Constraints}
As discussed in Section~\ref{sec:flex_modeling}, $(p,q)$ are constrained by generator, current, and voltage constraints. To deploy linear optimization techniques, such as displaying the feasible set as a polytope, all constraints must be linearized.
Branch current magnitudes are constrained to ensure that thermal line limits are not exceeded, which is enforced by:
\begin{equation} \label{eq:Ilim}
    \|I_{\mathrm{init},l,m}+\Delta I_{l,m}\| \leq I^\mathrm{max}_{l,m}, \qquad \forall (l,m)\in\mathcal{E}.
\end{equation}
This quadratic current constraint is linearized using a piece-wise linear approximation from \cite{Yang2016}. Similarly, the nonlinearities in generator constraints $\mathcal{X}_i,\forall i\in\mathcal{C}$ are linearized using the same approach. In general, the more linear constraints are used to approximate a nonlinear region, the more accurate the approximation but the dimensionality of the problem increases. 

To avoid introducing further computational complexity, voltage deviation is not piece-wisely approximated. Instead, under the assumption that the angle of the voltages is small, only the real component of the voltage deviation is constrained: 
\begin{equation} \label{eq:Vlim}
	V^\mathrm{min}_i \leq \mathfrak{R}(V_{\mathrm{init},i}+\Delta V_i) \leq V_i^\mathrm{max}, \quad\forall i\in\mathcal{N}.
\end{equation}
%where $V_i^\mathrm{min}\in\R_{\geq 0}$ and $V_i^\mathrm{max}\in\R_{\geq 0}$ are the minimum and maximum allowed voltage magnitudes, defined for each node $i\in\mathcal{N}$. 
Finally, by substituting \eqref{eq:BFS_final} into \eqref{eq:Ilim} and \eqref{eq:Vlim}, and enforcing generator constraints $(p_i,q_i)\in\mathcal{X}_i,\forall i\in\mathcal{C}$, we obtain a set of linear inequalities $\bar{\mathcal{F}}\coloneqq\{x\in\R^{n_x}\mid Ax\leq b\}$ approximating the feasible set given in \eqref{eq:feasible_space}. Here, $x\coloneqq(\Delta P, \Delta Q, \Delta p,\Delta q)$ is the vector of decision variables with $n_x\coloneqq 2n_c+2$, and matrices $A\in\R^{m\times n_x}$ and $b\in\R^{m\times1}$ contain coefficients of the considered $m\in\N$ inequalities. Note that the number of inequalities $m$ depends not only on the size of the network, but also on the number of linear constraints used to approximate a nonlinear inequality constraint, as discussed in \eqref{eq:Ilim}.

\subsection{Fourier-Motzkin Elimination} \label{sec:FMEtheory}
To compute the network flexibility on the $PQ$-plane, the polytope defined by $\bar{\mathcal{F}}$ needs to be reduced to and displayed in terms of $(\Delta P,\Delta Q)$. To this end, the Fourier-Motzkin Elimination method \cite{fouriermotzkin} for solving a set of linear inequalities is used. The FME works by eliminating one decision variable from the constraint set $Ax\leq b$ at a time. Without the loss of generality, we can reformulate the set $\bar{\mathcal{F}}$ to isolate any variable to be eliminated, e.g., let us select $x_{n_x}$ from $x\coloneqq(x_1,\dots, x_{n_x})$ and rearrange the inequalities as follows:
\begin{align}
    A_i^\prime x^\prime &\leq b_i^\prime, &&\forall i\in\mathcal{I},\\
    A_j^\prime x^\prime - x_{n_x} &\leq b_j^\prime \iff x_{n_x}\geq A_j^\prime x^\prime - b_j^\prime,  &&\forall j \in\mathcal{J}, \label{eq:neg_x}\\
    A_k^\prime x^\prime + x_{n_x} &\leq b_k^\prime\iff x_{n_x} \leq b_k^\prime- A_k^\prime x^\prime, &&\forall k\in\mathcal{K}, \label{eq:pos_x}
\end{align}
where $x^\prime\coloneqq(x_1,\dots,x_{n_x-1})$ is the vector of decision variables without $x_{n_x}$, $A_i^\prime$ is a submatrix of $A$ with the column corresponding to $x_{n_x}$ removed, and $b_i^\prime$ is a subvector of $b$ with the row corresponding to $x_{n_x}$ removed. Sets $\mathcal{I},\mathcal{J},\mathcal{K}$ contain indices of rows in $A^\prime,b^\prime$ related to constraints in which $x_{n_x}$ does not appear, appears with a negative sign, and appears with a positive sign, respectively. By rearranging and combining \eqref{eq:neg_x} and \eqref{eq:pos_x}, we can eliminate the variable $x_{n_x}$ and obtain a reduced linear inequality system:
\begin{align}
    A_i^\prime x^\prime &\leq b_i^\prime, &&\forall i\in\mathcal{I} \\
    A_j^\prime x^\prime - b_j^\prime &\leq x_{n_x} \leq b_k^\prime- A_k^\prime x^\prime, &&\forall j\in\mathcal{J},\forall k\in\mathcal{K}. \label{eq:fme_reduced}
\end{align}
Notice that in \eqref{eq:fme_reduced} variable $x_{n_x}$ can be dropped, and the new inequality writes as $A_j^\prime x^\prime - b_j^\prime \leq b_k^\prime - A_k^\prime x^\prime$. The new set of inequalities has the same solutions as the initial set over the remaining variables in $x\prime$ \cite{fouriermotzkin}.
The FME procedure is repeated until all variables pertaining to DERs are removed, which results in the same solution space as the initial system $\bar{\mathcal{F}}$ over the remaining variables $(\Delta P,\Delta Q)$.
This constraint set corresponds to the desired FOR on the $PQ$-plane, which represents the aggregated flexibility of the ADN's DERs under consideration of network and generator constraints.

\section{Addressing Uncertainty in DER Generation Through Chance Constraints} \label{sec:CCO}
Most distributed resources are subject to uncertain deviations in power output due to continuous changes in wind conditions, solar irradiation, or load fluctuations. Hence, when deploying DERs as a source of flexibility, this uncertainty should be taken into account. In this work, Chance-Constraints (CCs) \cite{roald} are used to allow system operators to adjust the level of conservatism in the FOR model. 
We address the uncertainty associated with active power output of PV; however, flexible loads, wind turbines and other DERs could be included without any conceptual changes.
To this end, we model the uncertain active power outputs of PVs as $\Tilde{p}_i = p_i + \omega_i$, where $p_i\in\R_{\geq0}$ is the forecasted PV output and $\omega_i\in\R$ are the zero mean fluctuations for each unit $i\in\mathcal{P}$.

The chance-constrained expression for the PV power output can now be formulated as follows: 
\begin{align} \label{eq:cc_pv}
    \mathbb{P}\Big\{(\Tilde{p}_i(\omega_i),q_i)\in\mathcal{X}_i\Big\}\leq1-\varepsilon,\quad\forall i\in \mathcal{P}, 
\end{align}
where the operator $\mathbb{P}\{\cdot\}$ indicates a transformation of the inequality constraint into a chance constraint, and $1-\varepsilon$ is the desired constraint satisfaction probability. 
Note that this CC is intractable as it is semi-infinite, due to the uncountable number of possible realizations of the uncertain variable $\omega$. Approximations of CC problems have been widely explored in literature. We employ two approximations: the the user-friendly Quantile Method (QM) \cite{roald} and the well-known Scenario Approach (SA) \cite{campiSP}, and compare their performances.

\subsection{Quantile Method} \label{sec:QM}
In \cite{roald}, it was found that the chance constraints can be reformulated by introducing uncertainty margins $\lambda_i\in[0,p_i^\mathrm{max}]$ to tighten the original constraint, as follows:
\begin{align} \label{eq:cc_reformulation}
     \Tilde{p}_i = p_i + \lambda_i,\quad \forall i\in\mathcal{P}.
\end{align}
To determine the values of $\lambda_i$, we first obtain the probability distribution of each PV units' power output at the location, date and time of interest. Then, we compute the upper quantile $(1-\varepsilon)$ of each distribution and set the uncertainty margins for the PV units as $\forall i \in \mathcal{P}$ as $\lambda_i = p^{1-\varepsilon}_i- p_i^\mathrm{pv}$. Here, $p^{1-\varepsilon}_i\in\R$ is a PV output data point that is greater than $(1-\varepsilon)$ of the rest of the PV data observed at that time instant.

\subsection{Scenario Approach} \label{sec:scenarioapp}
In the scenario approach \cite{campi2008}, the CC problem is approximated by randomly picking $N_s\in\N$ scenarios from the uncertain variable's probability distribution and ensuring that constraints are satisfied for all drawn samples $\{\xi_j\}_{j=1}^{N_s}$. To determine the minimum number of samples that must be drawn such that the chance constraints are not violated for $1-\varepsilon$ of the probability mass with a confidence of $1-\beta>0$, the equation from \cite{GENG2019341} is rewritten for one decision variable ($p_i$), as:
\begin{equation}\label{eq:solveforN}
    N_s = \ln{(\beta)}/\ln{(1-\varepsilon)},
\end{equation}
where violation probability $\varepsilon$ and confidence parameter $\beta$ are chosen depending on the level of desired conservatism. 
We apply the scenario approach by drawing $N_s$ samples from the distribution of each PV unit's recorded power output at a given location, time, and day of the year for multiple years. Considering that the lowest drawn power output sample $p_{s,i}^\mathrm{min}=\min_{j\in\{1:N_s\}}{\xi_{i,j}}$ tightens the PV capability curve the most, it is used to compute the uncertainty margin as: $\lambda_i=p_{s,i}^\mathrm{min}-p_i^\mathrm{max}$. Thus, the same CC approximation \eqref{eq:cc_reformulation} is used, however, with differently computed uncertainty margin.

\section{Results} \label{sec:res}
\begin{figure}[!t]
    \centering
    \includegraphics[scale=0.525]{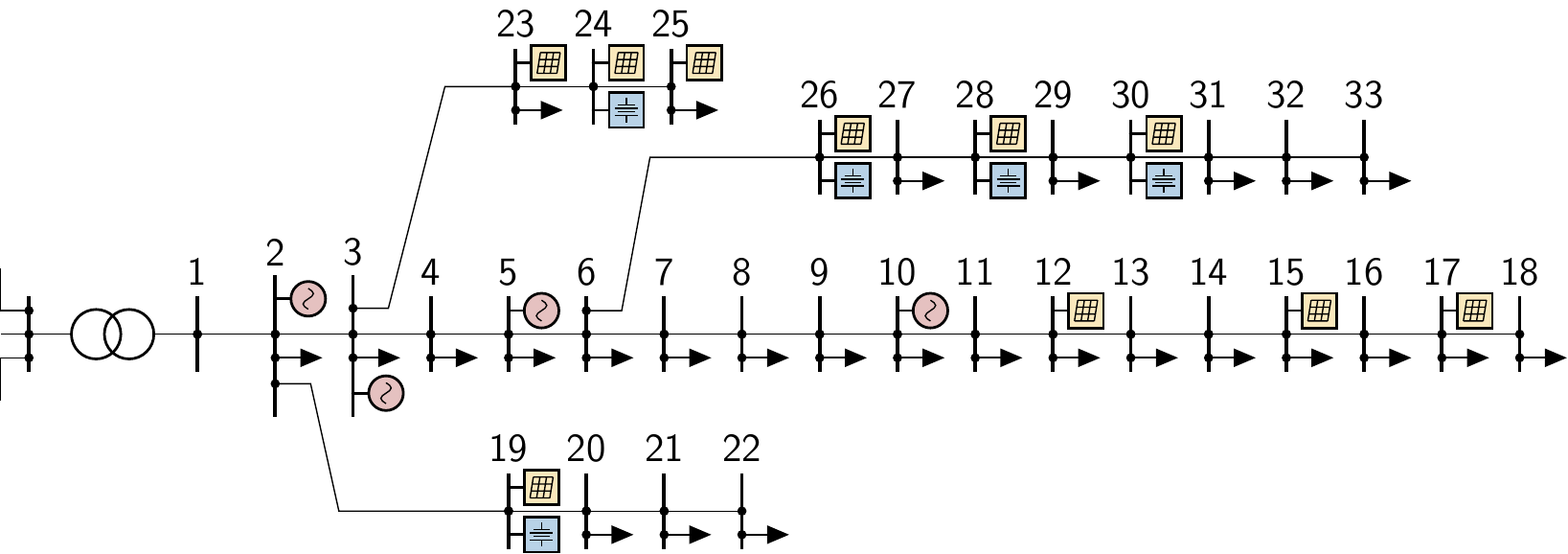}
    \caption{Modified IEEE 33-bus network with a high density of DERs.}
    \label{fig:33busIEEE}
    \vspace{-0.35cm}
\end{figure}

The proposed FOR mapping method has been tested and evaluated on a modified version of the IEEE 33-bus network \cite{Riaz2019}, presented in Fig.~\ref{fig:33busIEEE}. The system has been modified by adding PV units at nodes $\mathcal{P}=\{12,15,17,19,23,24,25,26,28,30\}$, storage at nodes $\mathcal{B}=\{19,24,26,28,30\}$ and DG units at nodes $\mathcal{D}=\{2,3,5,10\}$. The total active power capacity of the four diesel generators is \SI{1}{\mega\watt}, with individual generator limits between \SI{100}{\kilo\watt} and \SI{500}{\kilo\watt}. Each of the ten PV units have a real power capacity between \SI{28}{\kilo\watt} and \SI{38}{\kilo\watt}, amounting to the total installed capacity of \SI{353}{\kilo\watt}. It is assumed that $a = 0.9$. Lastly, \SI{350}{\kilo\watt} of BESS capacity are installed, with each unit’s real power limits between \SI{50}{\kilo\watt} and \SI{100}{\kilo\watt}. The minimum and maximum acceptable voltages at each bus are set to $0.9$ and $1.1$ p.u., respectively, and the thermal loading limits of each line are set to $125\%$ of the current magnitude corresponding to the base load and DER injections. YALMIP and MPT3 toolboxes \cite{yalmip,mpt3} are used for implementation and execution of the considered FOR mapping methods.

\subsection{Comparative Analysis}
In this section, the proposed approach (denoted as FME FOR) is compared against the GSK-based FOR mapping method from \cite{Fortenbacher2020}, and an optimization approach presented in \cite{Silva2018} is used as a benchmark. The key idea of the optimization method is to iteratively solve a nonlinear OPF problem to outline the borders of the FOR. This is done by fixing the reactive power flow across the PCC at a chosen value $Q_0$ and finding the maximum feasible active power flow $P_0^\mathrm{max}$ across the PCC, corresponding to that $Q_0$ value. The procedure is repeated until all boundary $(P_0^\mathrm{max},Q_0)$ tuples are identified. The GSK-based FOR method from \cite{Fortenbacher2020} is a geometric method that displays the feasible region of linear current, voltage and generator constraints. Its key difference to our method is that the relationship between individual generator power injections and the power flow across the PCC is directly related using GSKs instead of linear power flow and FME. 

Two key metrics are used to measure the relative accuracy of the selected methods: the fill factor $\phi\in[0,1]$ and the error $\delta\in[0,1]$ defined by
\begin{equation}
    \phi = \frac{A^\star\cap A}{A^\star},\quad\delta = 1-\frac{A^\star\cap A}{A},
\end{equation}
where $A$ represents the approximated polyhedral area, and $A^\star$ denotes the reference area based on the optimization method.   

\begin{figure}[!t]
    \centering
    \includegraphics[scale=1]{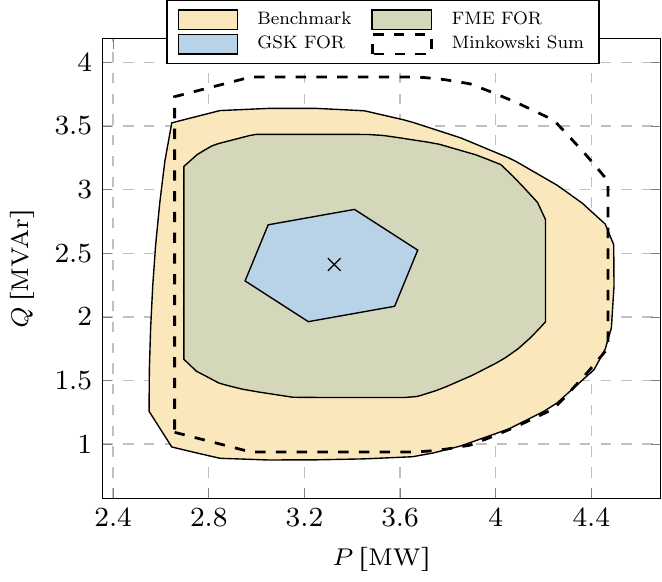}
    \caption{Feasible operating region of the modified IEEE 33-bus system computed using: (i) the novel FME FOR approach; (ii) the GSK-based approach \cite{Fortenbacher2020}; (iii) the benchmark optimization approach \cite{Silva2018}; and (iv) the Minkowski sum method \cite{Zhao20174721}. Symbol $\times$ denotes the initial operating point.}
    \label{fig:FOR_Comparison}
    \vspace{-0.35cm}
\end{figure}

The feasible operating region of the modified 33-bus system computed using the aforementioned methods is presented in Fig.~\ref{fig:FOR_Comparison}. The Minkowski sum of the individual generator polytopes is also presented to show the network-agnostic flexibility potential. It does not consider current and voltage constraints, thus deemed inadequate to estimate the flexibility of the network, as displaying flexibility significantly beyond the true FOR may cause dangerous scheduling conflicts. As can be seen from the figure, the FME FOR method reveals more of the flexibility compared to the GSK-based FOR mapping. The fill factors for the two methods, with the reference area corresponding to the feasible space computed using the benchmark method, are $0.61$ and $0.10$ respectively. Further comparisons for different network configurations are presented in Table~\ref{tab:config_comparison}. It is evident that for all considered configurations the proposed method outperforms the previously developed GSK-based FOR mapping method. The use of GSKs limits the FOR computation to always result in interdependent, proportional generator injection scenarios, which are unlikely to reflect a network's full potential. The proposed FME-based method omits GSKs and therefore proves to be more consistent across networks, not needing tailoring of GSKs to network configurations.
However, due to linearization errors introduced by the linear power flow approximation, the proposed method introduces a small error in the FOR in some cases. Further, we observe that fill factors are smaller for networks in which large generation is installed far from the PCC, since the binding piece-wise linear approximations of current constraints are under-approximations which accumulate along lines, hence tightening current limits beyond required bounds.
\begin{table}[!b]
\vspace{-0.35cm}
\renewcommand{\arraystretch}{1}
\caption{Comparison of fill factors and errors for different network configurations.}
\label{tab:config_comparison}
\noindent
\centering
    \begin{minipage}{\linewidth} %Use the minipage environment to footnote tables
    \begin{center}
        % values in brackets: c (column is centered), l (column is left aligned)
\scalebox{0.85}{%
    {\setlength{\extrarowheight}{.15em}\tabcolsep=4pt
    \begin{tabular}{c||c c c | c c|c c}
    \toprule
    \multirow{2}{*}{Config.} & \multirow{2}{*}{$\mathcal{P}$} & \multirow{2}{*}{$\mathcal{B}$} & \multirow{2}{*}{$\mathcal{D}$} &\multicolumn{2}{c}{FME FOR} & \multicolumn{2}{c}{GSK FOR} \\
            \cline{5-8}
             & & & & $\phi$ & $\delta$ & $\phi$ & $\delta$  \\
    \hline
    1 & \begin{tabular}{@{}c@{}}33,29,28,26,24,\\23,22,18,12,11 \end{tabular} & 33,29,28,26,24 & 2,3,4,7 & 0.78 & 0.02 & 0.64 & 0\\
    \arrayrulecolor{black!30}\cline{1-4}
    2 & \begin{tabular}{@{}c@{}}31,28,27,26,25,\\22,16,14,9,8 \end{tabular} &31,28,27,26,25 & 2,3,4,7 & 0.77 & 0.01 & 0.67 & 0\\
    \arrayrulecolor{black!30}\cline{1-4}
    3 & \begin{tabular}{@{}c@{}}5,24,21,6,25,\\22,7,26,23,8 \end{tabular} & 5,24,21,6,25 & 2,3,4,20 & 0.75 & 0.08 & 0.25 & 0\\
        \arrayrulecolor{black!30}\cline{1-4}
    4 & \begin{tabular}{@{}c@{}} 5,6,8,9,10,\\11,12,13,14,15 \end{tabular} & 5,6,8,9,10 & 2,23,4,7 & 0.55 & 0 & 0.05 & 0\\
    \arrayrulecolor{black}\bottomrule
    \end{tabular}}
    }
    \end{center}
    \end{minipage}
\end{table}

%In Table~\ref{tab:config_comparison} it can be seen that the fill factor of the novel FME-based method is more consistent across networks networks than those of the GSK-based method. Hence, the quality of the FME-based method does not depend on the case study network which it is tested on as much as the GSK-based method does. Though not shown here, the GSK-based method's fill factor strongly depends upon the GSK type definition (maximum-capacity based, equality-based, ...) and the optimal GSK definition also varies with network. The proposed FME-based method omits GSKs and was found to be more consistent across networks, not require tailoring of GSKs to network configurations, and obtain higher fill-factors than the GSK-based method throughout.

The computational performance is evaluated by measuring the CPU time for the algorithm execution, averaged over $100$ runs. The computation time measured for the proposed method is \SI{0.79}{\second}, \SI{10}{\milli\second} for the GSK-based FOR and \SI{7.5}{\second} for the benchmark method. Therefore, the proposed method has computational advantages over the benchmark method. Such rapid FOR approximations may be required for ancillary services that operate at short time-scales and cannot afford the optimization-based method's computation time. 
%Hence, despite the FOR FME being an approximation of the true FOR, it could be a relevant tool for system operators. In particular, the Fast FOR FME shows the flexibility around the setpoints very well, and only far-away points are not displayed well. Due to ramping times associated with reaching operating points far from the setpoint, flexibility near the setpoint is more likely to be deployed on short time-scales, making the fast FOR mapping method a relevant tool for rapid approximations. 
% Therefore, the proposed method has a computational advantages over the benchmark method while being more accurate than the previously proposed fast mapping method.

% The limitations of the proposed model: 
% Moreover, linear piece-wise approximation of generator and current constraints, introduces a trade-off between the computation time of the FME procedure and the accuracy of the linear approximation of the constraints. 

% It was found that current constraints are the binding constraints in these configurations, followed by generator and voltage constraints.

\subsection{The Impact of PV Uncertainty}

\begin{figure}[!t]
    \centering
    \includegraphics[scale=1]{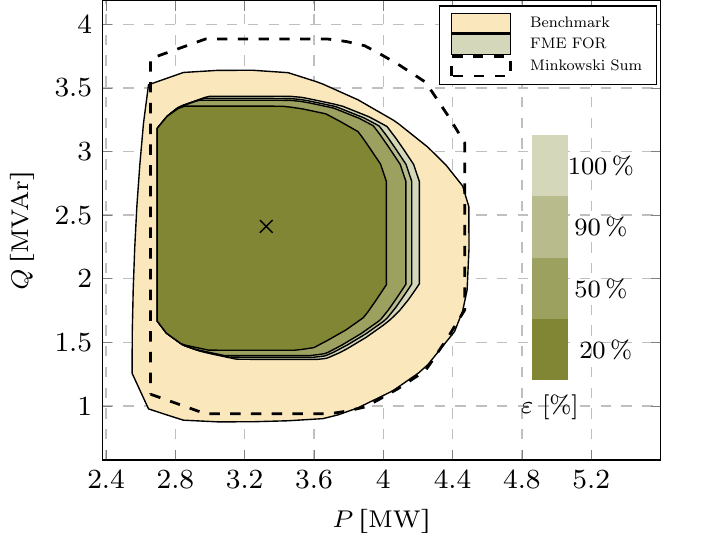}
    \caption{The impact of PV uncertainty on the FOR of the modified IEEE 33-bus system displayed using the quantile method for different $\varepsilon$ values.}
    \label{fig:FOR_Uncertainty_Impact}
    \vspace{-0.35cm}
\end{figure}

To demonstrate how the PV uncertainty impacts the FOR through chance-constraints of PV active power \eqref{eq:cc_pv}, we employ PV data from \cite{solarPV} 
%for the $22^\mathrm{nd}$ of July, 2020, at 14:10 
to create a distribution of the PV output. Figure~\ref{fig:FOR_Uncertainty_Impact} shows the FOR computed using the proposed method when the active power output of each PV unit is chance-constrained using the quantile method presented in Sec~\ref{sec:QM} and the confidence parameter $\varepsilon$ is varied. Low $\varepsilon$ values tighten the PV power output limits and thus lead to reduced FOR, while larger $\varepsilon$ values result in less conservative limits and larger FOR. Hence, by adjusting $\varepsilon$, operators are able to plan under varying levels of acceptable risks. 
%The quantile method captures the uncertainty well, as it is not too conservative and gives consistent results. 
% For example, a PV unit rated 38kW with the probability distribution calculated based on data from \cite{solarPV} is used for to determine the chance-constrained PV output for the 22nd of July, at 14:10 at a chosen violation probability of $\varepsilon = 0.90$. 
% It was found that in $90\%$ of cases on that day and time, the PV unit's active power output is limited to or below 30.8kW and for only $10\%$ of cases the active power limit lies above 30.8kW. Hence, the PV unit's output limit is updated to 30.8kW under this chance constraint. This is a relatively risky case.
% We apply this method to all PV units in the system and obtain a reduced FOR region, as shown in Fig.~\ref{fig:FOR_Uncertainty_Impact} as $\varepsilon = 0.90$ Fast FOR FME. 

Table~\ref{tab:scenario_approach} shows the fill factors that were found running the FME FOR with the scenario-based CC approximation, four times for each $\varepsilon$ and $\beta$ combination.
%The confidence parameters in the table can be interpreted as follows: e.g. for $\varepsilon = 0.9, \beta = 0.95$, the solution to the SP will hold for at least $1-\varepsilon = 10\%$ of the probability mass with a certainty of at least $1-\beta = 5\%$. This is a relatively risky case. 
The key drawback of the scenario approach becomes immediately apparent: for the same network and the same $\varepsilon$ and $\beta$ values, significantly different fill factors are found each time the algorithm is run. This behavior results from the fact that $N_s$ samples are drawn from the distribution to approximate the chance-constraint, and the smallest sample drawn ultimately becomes the new constraint on the PV output. Considering that the samples are drawn randomly, the smallest drawn sample may differ widely among runs. The advantage that the scenario and quantile methods share is that they are both data driven and do not significantly add to the computation time. However, the quantile approximation leads to consistent, reproducible and less conservative results and is more intuitive.
\begin{table}[!b]
\vspace{-0.4cm}
\renewcommand{\arraystretch}{1}
\caption{Scenario approach performance for multiple runs and different confidence parameters.}
\label{tab:scenario_approach}
\noindent
\centering
    \begin{minipage}{\linewidth} %Use the minipage environment to footnote tables
    \begin{center}
        % values in brackets: c (column is centered), l (column is left aligned)
\scalebox{0.9}{%
    {\setlength{\extrarowheight}{.3em}\tabcolsep=5pt
    \begin{tabular}{c c||c c c c}
    \toprule
    \multicolumn{2}{c}{Confidence parameters} & \multicolumn{4}{c}{Fill factors for different runs} \\
            \cline{1-6}
        $\varepsilon$ & $\beta$ & $\phi_1$ & $\phi_2$ & $\phi_3$ & $\phi_4$  \\
    \hline
    0.99 & 0.99 & 0.50 & 0.51 & 0.57 & 0.58 \\
    0.99 & 0.95 & 0.51 & 0.56 & 0.57 & 0.57 \\
    0.95 & 0.99 & 0.52 & 0.57 & 0.57 & 0.57 \\
    0.95 & 0.95 & 0.51 & 0.55 & 0.51 & 0.54 \\
    0.90 & 0.99 & 0.57 & 0.56 & 0.50 & 0.56 \\
    0.90 & 0.95 & 0.51 & 0.57 & 0.50 & 0.55 \\
    %\arrayrulecolor{black!30}\cline{1-4}
    \arrayrulecolor{black}\bottomrule
    \end{tabular}}
    }
    \end{center}
    \end{minipage}
\end{table}

\section{Conclusion} \label{sec:concl}
% \vspace{-0.05cm}
This paper presents a fast method to compute a linear approximation of the FOR at the TSO-DSO interface of radial ADNs. Our analysis of a modified IEEE 33-bus test system revealed that the novel FOR mapping method takes less than one second to compute and is about ten times faster than a benchmark optimization-based method, while finding a significantly more accurate approximation of an ADN's FOR than a previous fast mapping method. Secondly, it is shown that the quantile-based uncertainty model successfully captures the uncertainty while giving consistent, reasonably conservative results. The method is intuitive and allows the system operator to adjust the level of risk they are willing to take. The scenario approach was also tested but was found to be overconservative and inconsistent due to its random scenario selection step. In future work, we plan to expand the model to larger network sizes by computing the FOR for each feeder in parallel and then aggregating the individual feeders.

\bibliographystyle{IEEEtran}
% \vspace{-0.15cm}
\bibliography{bibliography}
% \vspace{-0.2cm}
% That's all folks
\end{document}